\documentclass{article}
\usepackage{spconf,amsmath,graphicx}
\usepackage{url, hyperref}
\usepackage{multicol}
\usepackage{tablefootnote}
\usepackage{enumitem}
\usepackage{multirow}
\usepackage{booktabs}

\title{StoryTTS: A Highly Expressive Text-to-Speech Dataset with Rich Textual Expressiveness Annotations}
\name{Sen Liu, Yiwei Guo, Xie Chen, $^{\dagger}$Kai Yu
\thanks{$^{\dagger}$Kai Yu is the corresponding author.}}
\address{MoE Key Lab of Artificial Intelligence, AI Institute\\ X-LANCE Lab, Department of Computer Science and Engineering\\Shanghai Jiao Tong University, Shanghai, China\\
\small{\texttt{\{sen.liu, yiwei.guo, chenxie95, kai.yu\}@sjtu.edu.cn}}
}

\begin{document}
\ninept
\maketitle
\begin{abstract}
While acoustic expressiveness has long been studied in expressive text-to-speech~(ETTS), the inherent expressiveness in text lacks sufficient attention, especially for ETTS of artistic works. In this paper, we introduce StoryTTS, a highly ETTS dataset that contains rich expressiveness both in acoustic and textual perspective, from the recording of a Mandarin storytelling show.
A systematic and comprehensive labeling framework is proposed for textual expressiveness. We analyze and define speech-related textual expressiveness in StoryTTS to include five distinct dimensions through linguistics, rhetoric, etc. Then we employ large language models and prompt them with a few manual annotation examples for batch annotation.
The resulting corpus contains 61 hours of consecutive and highly prosodic speech equipped with accurate text transcriptions and rich textual expressiveness annotations.
Therefore, StoryTTS can aid future ETTS research to fully mine the abundant intrinsic textual and acoustic features. Experiments are conducted to validate that TTS models can generate speech with improved expressiveness when integrating with the annotated textual labels in StoryTTS. 
\end{abstract}


\section{Introduction}
\label{sec:intro}
The advancement of deep learning has significantly enhanced the quality of text-to-speech~(TTS), enabling TTS models~\cite{tacotron,tacotron2,FastSpeech2,vits,RichProsody,ywg1} to produce speech that closely resembles human speech. However, these models tend to excel in synthesizing speech with relatively simple emotional characteristics. 
When it comes to expressive performance genres such as novels, poems, talk shows, and others, these methods still fall short of delivering the desired level of expressiveness. 
Such performances are usually rich in expressive text.
This form of text influences the cadence and rhythm of the speaker's delivery and refers to a form of written language that can bring to life the meaning and feeling of the subject matter to be conveyed.
A common approach is mine semantic and syntactic features from the expressive text. 
\cite{paper1} introduced a TTS model that explicitly incorporates text-context semantic information extracted from a pre-trained BERT~\cite{bert} model. This approach effectively enhances the expressiveness of the synthesized speech. 
In \cite{paper2}, a syntax graph is constructed for each input sentence based on its dependency tree. Following this, they employed a graph encoder to extract syntactic features, resulting in improved duration and pitch prediction.

However, these studies often rely on the coarse-grained semantic representations of pre-trained language models or basic syntactic structures, and they have not conducted a comprehensive and thorough exploration of textual expressiveness, especially speech-related expressiveness.
To achieve natural and expressive TTS synthesis, it's essential to convey the emotional stance of the text, which requires identifying speech-related textual expressive features in the text.
This is also supported by literature~\cite{readexpre}, where a thorough investigation was conducted into the relationship between prosody and linguistics. This study underscored that elements like the emotional content of the text, sentence patterns, and syntax have a direct impact on reading expressiveness.
Hence, understanding how to generalize, summarize, and characterize speech-related expressive features from expressive text might be crucial for expressive TTS.

In this paper, we introduce StoryTTS, a highly expressive TTS dataset with rich expressiveness from both acoustic and textual perspectives.
We initially construct the dataset from the recording of a Mandarin storytelling show with careful revision of transcripts and punctuations.
Then we establish a systematic and comprehensive labeling framework for textual expressiveness.  
Specifically, we analyze and define speech-related textual expressiveness in StoryTTS to include five distinct dimensions through linguistics, rhetoric, and literary studies. These dimensions include rhetorical devices, sentence patterns, scenes, imitated characters, and emotional colors.
Then we employ large language models~(LLMs) and prompt them with a few manual annotation examples for batch annotation. The resulting corpus contains 61 hours of consecutive and highly prosodic speech equipped with accurate text transcriptions and rich textual expressiveness annotations.
We further conduct experiments to validate that TTS models can produce speech with enhanced expressiveness when integrating the annotations from StoryTTS.
Our contributions can be summarized as follows:
\begin{itemize}[leftmargin=*,noitemsep, topsep=5pt]
\setlength{\itemsep}{2pt}
    \item We construct StoryTTS, the first TTS dataset that contains rich expressiveness in both speech and texts and is also equipped with comprehensive annotations for speech-related textual expressiveness. This dataset is also of high acoustic quality, organized by consecutive chapters, and of sufficient size. We release the StoryTTS dataset online\footnote{\url{https://goarsenal.github.io/StoryTTS}}.
    \item We establish a framework powered by LLMs to annotate textual expressiveness in five distinct dimensions.
    \item We conduct experiments to validate that TTS models can produce speech with enhanced expressiveness when integrating the annotated textual expressiveness labels.
 \end{itemize}
\section{StoryTTS}
\vspace{-5pt}
The construction of StoryTTS is introduced in detail in this section, including data selection and retrieval, audio quality analysis, speech segmentation and automatic recognition, manual correction of recognition errors, and the enhancement of punctuation. The detailed statistics of StoryTTS are shown in Table \ref{table:dataset}.
\vspace{-12pt}
\begin{table}[h]
  \centering
  \caption{Detailed statistics of StoryTTS.}
  \vspace{3pt}
  \begin{resizebox}{0.9\columnwidth}{!}
  {
  \begin{tabular}{lcc}
    \toprule
    \textbf{Domain} & \textbf{Feature} & \textbf{Statistics}  \\
    \midrule
    \multirow{5}*{Speech} & Number of speakers & 1 \\
    & Total duration (hours)    & 60.9 \\
    & Mean duration per utterance (seconds) & 6.8 \\
    & Sampling rate (kHz) & 16 \\
    & Estimated SNR (dB) & 32 \\
    \midrule
    \multirow{2}*{Text} & Chapters & 160 \\
    & Sentences & 33108\\
    \bottomrule
  \end{tabular}
  }
  \end{resizebox}

  \label{table:dataset}
\end{table}
\vspace{-23pt}

\begin{table}[h]
  \centering
  \caption{Comparison of often used publicly available TTS datasets. Here, `\textbf{Lang}' refers to the respective language, `\textbf{Num of spks}' means the number of speakers, `\textbf{Hours / spk}' indicates the average hours per speaker, `\textbf{Pitch Std. / spk}' represents the average standard deviation of pitch per speaker, and `\textbf{Expr. Annots.}' denotes whether expressiveness annotations are available.}
  \vspace{3pt}
  \begin{resizebox}{1.0\columnwidth}{!}
  {
  \begin{tabular}{lccccc}
    \toprule
    \multirow{2}{*}{\textbf{Dataset}} & \multirow{2}{*}{\textbf{Lang}} & \textbf{Num of}  & \textbf{Hours} & \textbf{Pitch Std.} & \textbf{Expr.}\\
    ~ &  ~ & \textbf{spks} &  \textbf{/ spk} & \textbf{/ spk} & \textbf{Annots.} \\
    \midrule
    LJSpeech~\cite{ljspeech} & EN & 1 & 24 & 49.93 & N \\
    Blizzard-2013~\cite{BC2013} & EN & 1 & 319 & 49.32 & N \\
    HiFi-TTS~\cite{hifitts} & EN & 10 & 29.2 & 55.09 & N \\
    LibriTTS~\cite{LibriTTS} & EN & 2456& 4.2 & 48.51 & N \\
    Aishell3~\cite{aishell3} & ZH & 218 & 0.39 & 39.63 & N \\
    Biaobei~\cite{biaobei} & ZH & 1 & 12 & 58.57 & N \\
    \midrule
    StoryTTS & ZH & 1 & 60.9 & \textbf{98.22} & \textbf{Y} \\
    \bottomrule
  \end{tabular}
  }
  \end{resizebox}

  \label{table:compare}
\end{table}
\vspace{-18pt}
\subsection{Data selecting and retrieval}
\vspace{-3pt}
Storytelling, also known as ``Pingshu'', is a traditional Chinese oral art form where performers narrate stories, imitate various voices, and portray characters to enthrall audiences.
This form of oral art is usually based on historical novels, which makes storytelling not only rich in speech prosody but also diverse in textual expressiveness, such as linguistic structures, figures of speech, role-playing, etc.
Hence, storytelling shows satisfy our goal to a great extent.
We selected a storytelling show titled ``Zhi Sheng Dongfang Shuo'', recounting the legend of Dongfang Shuo, a key figure in the development of the Han Dynasty in ancient China. This performance is skillfully delivered by a female artist, Lian Liru.
To construct the dataset, we retrieved the recorded speech data from a public website, which is organized into 160 consecutive chapters.
These chapters have an approximate duration of 24 minutes each, hence amounting to 64 hours in total, including interval breaks.

\vspace{-8pt}
\subsection{Audio quality analysis}
\vspace{-3pt}
We also estimated the signal-to-noise ratio~(SNR) of the speech data, where the noise power was computed using the silence segments predicted by a voice activity detection~(VAD) tool.
As can be seen in Table \ref{table:dataset}, the SNR is estimated to be 32dB, indicating the high audio quality of the waveforms.
Subsequently, we conducted statistical analysis on multiple common Mandarin~(ZH) and English~(EN) datasets, as shown in Table \ref{table:compare}. The results reveal that StoryTTS exhibits a significantly higher pitch standard deviation than other datasets, providing compelling evidence of its substantial acoustic expressiveness.
Furthermore, StoryTTS includes expressiveness annotations, which will be further explained in the next section.
\vspace{-8pt}
\subsection{Speech segmentation and automatic recognition} 
\vspace{-3pt}
To process the originally coarsely segmented speech data, we implemented a three-step approach. We first employed a VAD tool to segment the chapter-level speech into utterances based on the duration of silence segments.
Long silences were also removed in this step, resulting in 60.9 hours of speech.
Subsequently, given the absence of matching text transcripts, we obtained text transcripts using Whisper~\cite{whisper}, a popular speech recognition model. 
We observed that there still exist speech segments that remained excessively long after the VAD process.
To address this issue, we identified these speech segments and their corresponding texts. 
We manually divided the prolonged text into smaller sentences and then utilized the Aeneas\footnote{\url{https://github.com/readbeyond/aeneas}} tool for synchronizing text fragments with speech. This alignment allowed us to accurately cut the speech, producing a final dataset of 33108 pairs of speech and text.

\vspace{-8pt}
\subsection{Manual correction of recognition errors} 
\vspace{-3pt}
Given the extremely variable pitch and speaking rate in the storytelling performances, the speech recognition results exhibited a higher error rate compared to standard speech. In response to this challenge, we meticulously reviewed every speech segment line by line and rectified the recognition errors. Furthermore, we have made diligent efforts to replace onomatopoeic elements in the speech with appropriate words from the corresponding text.
\vspace{-8pt}
\subsection{Punctuation enhancement}
\vspace{-3pt}
Punctuation plays a vital role in text presentation, conveying emotions like surprise or shock through exclamation points and indicating character dialogue or inner thoughts through double quotes. Although Whisper can identify some punctuation marks, it still falls far short of expectations. 
Thus, during our text review process, we made careful punctuation corrections and additions to ensure precise punctuation usage whenever possible. This attention to punctuation accuracy also significantly benefited our subsequent work on textual sentiment analysis.

\begin{figure*}[t]
  \centerline{\includegraphics[scale=0.43]{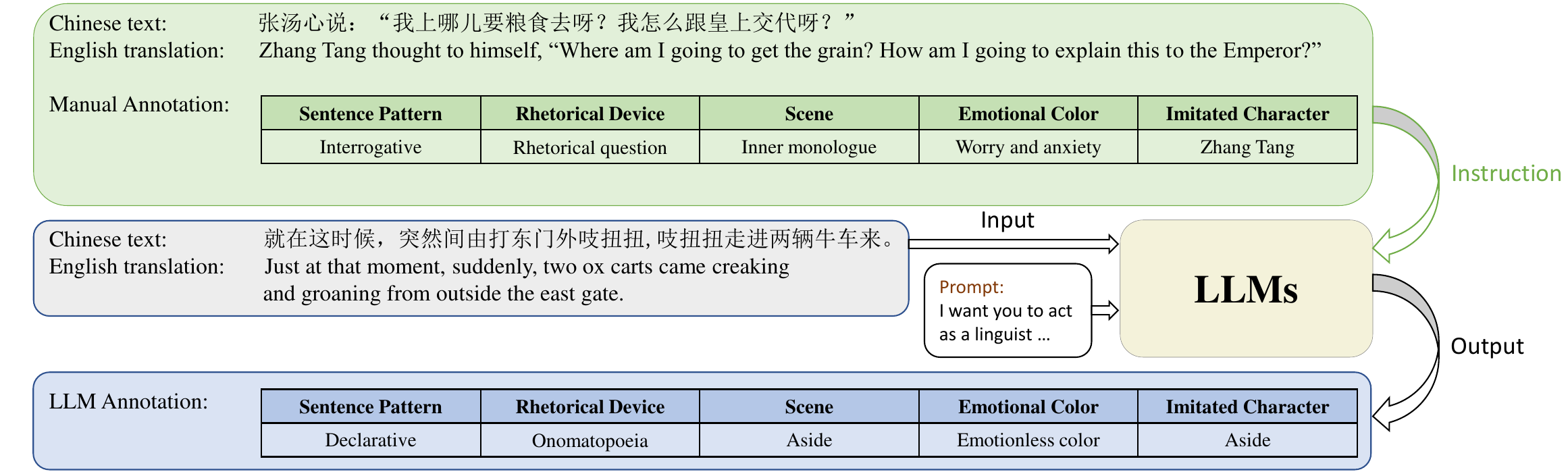}}
  \vspace{-10pt}
  \caption{An Illustration of Expressiveness Annotation Using LLMs.}
  \label{fig:text_example}
\end{figure*}	
\vspace{-5pt}
\section{LLM-driven expressiveness annotation}
\vspace{-5pt}
The expressive texts within StoryTTS exhibit a high degree of colloquialism and are rich in role-playing, psychological, action, and environmental descriptions. Inspired and guided by \cite{readexpre}, we introduced a systematic and comprehensive labeling framework that harnesses the power of LLMs. 
\vspace{-8pt}
\label{sec:rules}
\subsection{Exploring textual expressiveness}
\vspace{-3pt}
In our investigation of speech-related textual expressiveness, we classify it into five dimensions drawn from the fields of literary studies, linguistics, and rhetoric. 
These dimensions include rhetorical devices, sentence patterns, scenes, imitated characters, and emotional colors.

{\textbf{Rhetorical devices}}, like hyperbole, and {\textbf{sentence patterns}}, such as declarative sentences, are commonly employed textual expressive devices. For instance, using an exclamatory sentence or incorporating rhetorical devices like hyperbole can evoke emotions of excitement or surprise.
We also employed {\textbf{scenes}} such as role-playing, taking into account the characteristics of StoryTTS. For example, a role-playing scene often carries strong emotional content, while an aside typically lacks emotional elements. 
The specific categorization regarding sentence patterns, scenes, and rhetorical devices is shown in Table \ref{table:label_cate}. 

\begin{table}[h]
\centering
  \caption{Specific categories of Sentence Pattern, Scene, and Rhetoric Device in StoryTTS. Here, `\textbf{Num}' denotes the number of categories.}
  \vspace{3pt}
 \begin{resizebox}{0.9\columnwidth}{!}
  {
\begin{tabular}{lcc}
\toprule
\textbf{Dimension}                & \multicolumn{1}{l}{\textbf{Num}} & \multicolumn{1}{c}{\textbf{Category}}                      \\ \hline
\multirow{2}{*}{Sentence Pattern} & \multirow{2}{*}{\textbf{4}}      & Declarative,\ Interrogative              \\
             ~ & ~ & Imperative,\ Exclamatory                \\\midrule
\multirow{2}{*}{Scene}            & \multirow{2}{*}{\textbf{3}}      & Role-playing,\ Aside                     \\
              ~ & ~ & Inner monologue                        \\\midrule
\multirow{4}{*}{Rhetoric Device}  & \multirow{4}{*}{\textbf{11}}      & Hyperbole,\ Antithesis,\ Onomatopoeia  \\
              ~ & ~ & Simile,\ Personification,\ Quotation    \\
              ~ & ~ & Irony,\ Rhetorical question,\ Hypophora \\
              ~ & ~ & Rhetorical repetition,\ Transition         \\ \bottomrule
\end{tabular}
}
\end{resizebox}
  \label{table:label_cate}
\end{table}

{\textbf{Emotional colors}} often have a direct impact on the performer's expression. Instead of categorizing them into various polarities or predefined categories, we have chosen a more precise approach: summarizing the emotional color of a sentence using several words. This method can provide a more accurate description of the text's emotion compared to traditional categorization.

\begin{figure}[htb]
    \centerline{\includegraphics[scale=0.4]{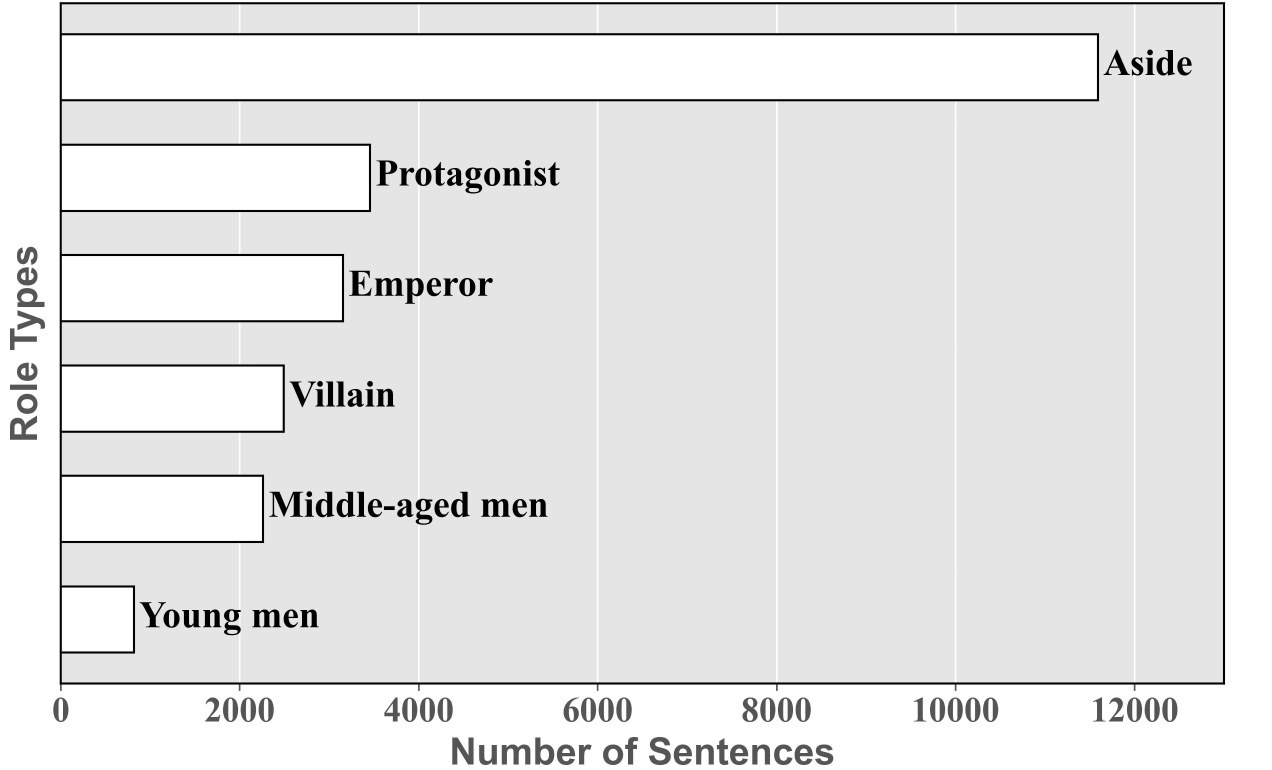}}
    \vspace{-10pt}
    \caption{Top 6 role types in StoryTTS.}
    \label{fig:role_type}
\end{figure}

In the case of {\textbf{imitated characters}} in StoryTTS, the performer often mimics the speech patterns of characters when delivering their lines. For example, she deliberately lowers the pitch and slows down when portraying an old man while raising the pitch and speeding up when mimicking a villain. We categorized the characters into \textbf{19} role types based on characteristics like age, gender, and status, which also served as the basis for the performer's mimicry. The six most frequent role types are shown in Figure \ref{fig:role_type}.
\vspace{-8pt}

\vspace{-10pt}
\subsection{Batch annotation via LLMs}
\vspace{-2pt}
LLMs have found applications in a wide range of studies. \cite{gpt3_annotator} demonstrated that GPT3~\cite{GPT3} performs well in data annotation tasks at a relatively low cost, making it a viable choice for individuals and organizations with limited budgets.  
To expedite the labeling process and reduce costs, we utilized GPT4~\cite{GPT4} and Claude2~\cite{Claude2}, both of which are more powerful than GPT3, for batch annotation.
During our annotation process, Claude2 was used for annotating sentence patterns, rhetorical devices, scenes, and imitated characters. However, when it came to summarizing emotional colors in the text, we found Claude2's performance to be suboptimal. Consequently, we turned to GPT4, which proved to be more proficient in this aspect.
In our prompt, we initiated the persona of a linguist for the LLMs. We then proceeded to guide the model, informing it that the input texts were consecutive and required labeling with contextual information. For instance, two consecutive sentences might belong to the same role-play scenario. Subsequently, we provided details about the sources and features of the text, emphasizing their richness in elements such as onomatopoeia, inner monologue, and role-plays.
Finally, we instructed the model to annotate each sentence in a prescribed format, adhering to the prompts and requirements outlined. These annotations align with the guidelines detailed in Section \ref{sec:rules}, where sentence patterns, scenes, rhetorical devices and imitated characters must be assigned specific categories, and each emotional color should be summarized in several words.

Initially, we attempted labeling in the zero-shot setting, but the results exhibited low accuracy. Consequently, we transitioned to the few-shot setting, where we provided the model with comprehensive and diverse labeled text and explained the rationale behind each labeling decision. In this setting, the model demonstrates improved accuracy and meets our labeling requirements. An illustration of annotation using LLMs is shown in Figure \ref{fig:text_example}.

\vspace{-10pt}
\section{Experiments and results}
\vspace{-4pt}
In this section, we build TTS models to analyze the impact of annotated textual expressiveness labels on synthetic speech.
\vspace{-7pt}
\subsection{Model Architecture}
\subsubsection{Baseline Model}
Our baseline model is implemented based on VQTTS~\cite{vqtts}, which utilizes self-supervised vector-quantized~(VQ)~\cite{vq-wav2vec,wav2vec2,XLSR53,HuBERT}, acoustic features rather than traditional mel-spectrogram. Specifically, it consists of an acoustic model, t2v, and a vocoder, v2w. T2v accepts the phoneme sequence and then outputs the VQ Acoustic feature and Auxiliary feature, which consist of pitch, energy, and probability of voice~\cite{POV}, and v2w receives them and thus synthesizes the waveform.

\vspace{-12pt}
\subsubsection{Expressiveness Encoder}
\vspace{-3pt}
To fully leverage our expressiveness annotations, we developed an expressiveness encoder. We employed four separate learnable embedding tables to supply information to the model for the four labels: sentence pattern, scene, rhetorical method, and imitated character. 
For each sentence, we assigned four category numbers based on these four expressive labels. We then inputted these numbers into the corresponding embedding tables, with vector dimensions of 32, 32, 64, and 256, respectively.

Regarding the modeling of emotional color, we employed distinct model structures. Given that emotion descriptions typically condense into several words, representing the overall sentiment of a sentence, while emotions may change within a single sentence. For instance, in an exclamatory sentence, emotion often intensifies towards the end. We initially extracted word-level embeddings for the entire sentence using a pre-trained BERT. Subsequently, we extracted the embedding of emotional color using a Sentence BERT. Through cross-attention~\cite{transformer} between these embeddings, we aimed to capture the distribution of emotions at different locations within the text, enhancing their expressive accuracy. Following this, we up-sampled the results to the phoneme level based on the word-to-phoneme correspondence and added them to the encoder output, along with the previous four embeddings.
\vspace{-11pt}
\subsection{Experimental setup}
\vspace{-3pt}
We conducted experiments to evaluate the impact of each of the five textual expressiveness labels on the expressiveness of the synthesized speech. Additionally, we assessed the cumulative effect of utilizing all these labels together.
For these experiments, we trained an acoustic model separately for 300 epochs using a batch size equal to 8. The vocoder was shared, and we trained 100 epochs on StoryTTS using a batch size of 8. The remaining model configurations and parameters remained consistent with those in \cite{dsetts}. Each experiment was performed on a single 2080 Ti GPU. 
To preprocess the text data, we utilized our internal Grapheme-to-Phoneme~(G2P) tool for text-to-phoneme conversion. We also set aside 5$\%$ of the text for test and validation sets, where the test set consists of 3 consecutive chapters. To obtain ground truth phoneme duration, we employed the Montreal Forced Aligner~\cite{MFA}, which conducts forced alignment using Kaldi~\cite{kaldi}.

\vspace{-10pt}
\subsection{Speech Synthesis Evaluation}
\vspace{-2pt}
\subsubsection{Metrics}
We performed a mean opinion score~(MOS) listening test involving 20 native listeners who were asked to rate each sample. MOS ratings were based on a 1-5 scale with 0.5-point increments and 95$\%$ confidence intervals. 
During our tests, we instructed listeners to specifically assess the level of expressiveness in the synthesized speech, all while evaluating speech quality.
For objective evaluations, we computed Mel-cepstral distortion~(MCD) using dynamic time warping~(DTW). Additionally, we analyzed the log F0 root mean square error~(log-F0 RMSE), also computed with DTW. MCD measures general speech quality, while log-F0 RMSE assesses performance in terms of speech prosody. Lower values for both of these metrics indicate better sound quality and rhythm in speech performance. 

\begin{table}[ht]
  \centering
    \caption{Results of evaluation for different setups. Here, `GT(Voc.) denotes the vocoded
ground truth speech, and `+ALL' means we utilized all the expressiveness labels together.}
    \vspace{3pt}
  \begin{resizebox}{1.0\columnwidth}{!}
  {
  \begin{tabular}{lccc}
    \toprule
    \textbf{Model} & \textbf{MCD} $\downarrow$ & \textbf{log-F0 RMSE} $\downarrow$ & \textbf{MOS} $\uparrow$  \\\midrule
    GT(Voc.) & 3.765 & 0.322 & $4.29\pm0.06$\\\midrule
    Baseline & 6.904 & 0.437 & $3.88\pm0.07$\\
    +Sentence Pattern & 6.746 & 0.432 & $3.90\pm0.08$\\
    +Scene & 6.692 & 0.431 & $3.90\pm0.07$\\
    +Rhetoric Device & 6.633 & 0.421 & $3.93\pm0.07$\\
    +Emotional Color & 6.271 & 0.412 & $3.92\pm0.08$\\
    +Imitated Character & 6.508 & 0.411 & $3.96\pm0.07$\\
    +ALL & \textbf{6.181} & $\textbf{0.402}$ & $\mathbf{4.09}\pm\mathbf{0.07}$\\
    \bottomrule
  \end{tabular}
  }
  \end{resizebox}
  \label{table:metric_result}
\end{table}
\vspace{-12pt}
\subsubsection{Results}
Table \ref{table:metric_result} presents the evaluation results. It can be seen that when incorporating the expressiveness labels into the model, both the subjective and objective scores outperform the baseline model. 
Specifically, sentence patterns and scene boosts were relatively minimal. This might be attributed to the prevalence of declarative sentences in the dataset, resulting in limited information acquired by the model. Additionally, while scene types are distributed fairly evenly, their diversity is insufficient to furnish the model with adequate information, owing to the different character imitations found in role-playing and inner monologue scenes.
Rhetorical devices and emotional colors bring more obvious enhancements.
Among the individual expressiveness labels, the imitated characters stand out as the most effective, as they directly provide information about the characters currently being mimicked, enabling the model to efficiently learn how the mimicked characters speak and thus synthesize speech close to that of the original data.

Finally, the fusion of all expressive labels provides the most significant enhancement. It outperforms other setups significantly in both objective and subjective metrics, supplying the model with increasingly accurate information about imitated characters and scenes. This fusion also benefits from the complementary nature of sentence patterns, rhetorical devices, and emotional colors.
\vspace{-8pt}
\section{conclusions}
This paper presented StoryTTS, the first TTS dataset that encompasses rich expressiveness from both acoustic and textual perspectives.  
Derived from a high-quality recording of a Mandarin storytelling show, this dataset serves as a valuable resource for researchers aiming to investigate acoustic expressiveness on the one hand.
Additionally, we conducted a comprehensive analysis of expressive text and categorized speech-related textual expressiveness into five distinct dimensions. Then we employed LLMs and provided them with a few manual annotation examples for batch annotation. The effective labeling of LLMs also offers insights for similar data labeling endeavors.
The dataset is thus equipped with abundant textual expressiveness annotations.
Experimental results demonstrated that TTS models can generate speech with significantly improved expressiveness when incorporated with the annotated textual expressiveness labels.
Future work may focus on integrating these expressiveness annotations with acoustic expressiveness to further enhance expressive speech synthesis.
\vspace{-8pt}
\section{Acknowledgements}
This work was supported by China NSFC Project (No.92370206), Shanghai Municipal Science and Technology Major Project \\
\noindent(2021SHZDZX0102) and the Key Research and Development Program of Jiangsu Province, China (No.BE2022059).

\newpage
\bibliographystyle{styles/IEEEbib}
\bibliography{citations/refs}
\end{document}